\documentclass[%
reprint,
11pt,
onecolumn,
 amsmath,amssymb,
 aps,
prb,
]{revtex4-2}

\usepackage{graphicx}
\usepackage{dcolumn}
\usepackage{bm}

\usepackage{lineno}
\usepackage{listings}
\usepackage{xcolor}

\definecolor{codegreen}{rgb}{0,0.6,0}
\definecolor{codegray}{rgb}{0.5,0.5,0.5}
\definecolor{codepurple}{rgb}{0.58,0,0.82}
\definecolor{backcolour}{rgb}{0.95,0.95,0.92}

\lstdefinestyle{mystyle}{
    backgroundcolor=\color{backcolour},   
    commentstyle=\color{codegreen},
    keywordstyle=\color{magenta},
    numberstyle=\tiny\color{codegray},
    stringstyle=\color{codepurple},
    basicstyle=\ttfamily\footnotesize,
    breakatwhitespace=false,         
    breaklines=true,                 
    captionpos=b,                    
    keepspaces=true,                 
    numbers=left,                    
    numbersep=5pt,                  
    showspaces=false,                
    showstringspaces=false,
    showtabs=false,                  
    tabsize=2
}

\usepackage{xurl}
\begin{document}


\title[]{Porting OpenACC to OpenMP on heterogeneous systems}

\author{Hicham Agueny}
\email{hicham.agueny@uib.no}
\affiliation{IT-Department,
University of Bergen, N-5007 Bergen, Norway \\
Norwegian Research Infrastructure Services (NRIS) 
}%

\begin{abstract}
This documentation is designed for beginners in Graphics Processing Unit (GPU)-programming and who want to get familiar with OpenACC and OpenMP offloading models. Here we present an overview of these two programming models as well as of the GPU-architectures. Specifically, we provide some insights into the functionality of these models and perform experiments involving different directives and discuss their performance. This is achieved through the use of a mini-application based on solving numerically the Laplace equation. Such experiments reveal the benefit of the use of GPU, which in our case manifests by an increase of the performance by almost a factor of 52. We further carry out a comparative study between the OpenACC and OpenMP models in the aim of converting OpenACC to OpenMP on heterogeneous systems. In this context, we present a short overview of the open-source OpenACC compiler Clacc, which is designed based on translating OpenACC to OpenMP in Clang. 

This documentation ultimately aims at initiating developers'/users' interest in GPU-programming. We therefore expect developers/users, by the end of this documentation, to be able to: 

\begin{itemize}
\item	Recognise the benefits of GPU-programming.
\item	Acquire some basic knowledge of the GPU-architecture and the functionality of the underlying models.
\item	Use appropriate constructs and clauses on either programming model to offload compute regions to a GPU device.
\item	Identify and assess differences and similarities between the OpenACC and OpenMP offload features.
\item Convert an OpenACC mini-application to OpenMP offloading.
\item	Get some highlights of available open-source OpenACC compilers.
\end{itemize}
\end{abstract}

\date{\today}

\maketitle

\tableofcontents

\section{Introduction}

OpenACC \cite{acc} and OpenMP \cite{omp} are the most widely used programming models for heterogeneous computing on modern HPC architectures. OpenACC was developed a decade ago and was designed for parallel programming of heterogenous systems (i.e. CPU host and GPU device). Whereas OpenMP is historically known to be directed to shared-memory multi-core programming, and only recently has provided support for heterogenous systems. OpenACC and OpenMP are directive-based programming models for offloading compute regions from CPU host to GPU devices. These models are referred to as Application Programming Interfaces (APIs), which here enable to communicate between two heterogenous systems and specifically enable offloading to target devices. The offloading process is controlled by a set of compiler directives, library runtime routines as well as environment variables. These components will be addressed in the following for both models with a special focus on directives and clauses. Furthermore, differences and similarities will be assessed in the aim of converting OpenACC to OpenMP.

$Motivation:$ NVIDIA-based Programming models are bounded by some barriers related to the GPU-architecture. 
Such models do not have direct support on different devices nor by 
the corresponding compilers. Removing such barriers is one of 
the bottleneck in GPU-programming, which is the case for instance with OpenACC. The latter is one of 
the most popular programming model that requires a special attention in terms of support on available architectures.  

As far as we know, the only compiler that fully supports OpenACC for offloading to both NVIDIA and AMD devices is GCC. The GCC's performance, however, suffers from some weaknesses and poses some challenges \cite{ieee}, which limit its extension. Although the Cray Compilation Environment (CCE) \cite{cee} has full support of OpenACC 2.0 and partial support of OpenACC 2.6, the support is limited only to Fortran, and thus no support for C or C++. This lack of support for OpenACC calls for an alternative that goes beyond the GCC compiler, and which ensures higher performance. On the other hand, the OpenMP offloading is supported on multiple devices by a set of compilers such as $Clang/Flang$ and $Cray$, and $Icx/Ifx$ which are well-known to provide the highest performance with respect to GCC. Therefore, converting OpenACC to OpenMP becomes a necessity to overcome the lack of stable implementations for all relevant hardware vendors, and to extend the OpenACC implementations to cover various GPU-architectures. In this context, there has been a project funded by the Exascale Computing Project \cite{project} conducted by J. Vetter et \textit{al.} \cite{ieee}, which aims at developing an open-source OpenACC compiler. This documentation is inspired by this work and is motivated by the need to document how to translate OpenACC to OpenMP on heterogeneous systems.

 This documentation is organised as follows. In sec. \ref{model}, we provide a computational model, which is based on solving the Laplace equation. Section \ref{acc2omp} is devoted to 
 the analysis of experiments performed using the OpenACC and OpenMP offload features and to a one-to-one mapping of these two models. Section \ref{clacccompiler} is directed to a discussion about open-source OpenACC compilers. Finally, conclusions are given in Sec. \ref{conclusion}.

\section{Computational model}\label{model}

We give a brief description of the numerical model used to solve the Laplace equation $\Delta f=0$. For the sake of simplicity, we solve the equation in a two-dimensional (2D) uniform grid according to
\begin{equation}\label{eq1}
\Delta f(x,y)=\frac{\partial^{2} f(x,y)}{\partial x^{2}} + \frac{\partial^{2} f(x,y)}{\partial y^{2}}=0. 
\end{equation}
Here we use the finite-difference method to approximate the partial derivative of the form $\frac{\partial^{2} f(x)}{\partial x^{2}}$. The spatial discretization in the second-order scheme can be written as 
\begin{equation}\label{eq2}
\frac{\partial^{2} f(x,y)}{\partial^{2} x}=\frac{f(x_{i+1},y) - 2f(x_{i},y) + f(x_{i-1},y)}{\Delta x}. 
\end{equation}
Inserting Eq. (\ref{eq2}) into Eq. (\ref{eq1}) leads to this final expression 
\begin{equation}\label{eq3}
f(x_i,y_j)=\frac{f(x_{i+1},y) + f(x_{i-1},y) + f(x,y_{i+1}) + f(x,y_{i-1})}{4}. 
\end{equation}

The Eq. (\ref{eq3}) can be solved iteratively by defining some initial conditions that reflect the geometry of the problem at-hand. This can be done either using Gauss?Seidel method or Jacobi method. Here, we apt for the Jacobi algorithm due to its simplicity. The corresponding compute code is written in Fortran 90 and is given below

\begin{lstlisting}[language=Fortran]
do while (max_err.gt.error.and.iter.le.max_iter) 
   do j=2,ny-1 
      do i=2,nx-1 
         df_x = f(i+1,j) + f(i-1,j) 
         df_y = f(i,j+1) + f(i,j-1) 
         f_k(i,j) = 0.25*(df_x + df_y) 
      enddo 
    enddo  
    max_err=0. 

    do j=2,ny-1 
       do i=2,nx-1 
          max_err = max(dabs(f_k(i,j) - f(i,j)),max_err) 
          f(i,j) = f_k(i,j) 
       enddo 
    enddo  
    iter = iter +1 
enddo 
\end{lstlisting}

\section{Comparative study: OpenACC versus OpenMP}\label{acc2omp}

In the following we first provide a short description of GPU accelerators and then perform experiments covering both the OpenACC and OpenMP implementations to accelerate the Jacobi algorithm in the aim of conducting a comparative experiment between the two programming models. The experiments are systematically performed with a fixed number of grid points (i.e. 8192 points in both $x$ and $y$ directions) and a fixed number of iterations that ensure the convergence of the algorithm. This is found to be 240 iterations resulting in an error of 0.001.

\begin{figure}[h!]
\centering
\includegraphics[width=16.cm,height=10cm]{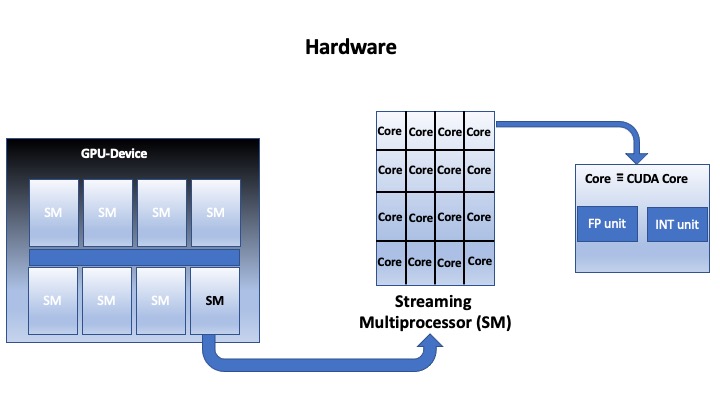}
\caption{\label{fig1} Schematic representation of a NVIDIA GPU-accelerator.}
\end{figure}

\subsection{GPU accelerators}\label{gpu}

We focus in this section on describing the NVIDIA GPU accelerator \cite{nvidia} as it is considered the most powerful accelerator in the world to be used for artificial intelligence (AI) and high-performance computing (HPC). The NVIDIA GPU-device consists of a block of a Streaming Multiprocessor (SM) each of which is organized as a matrix of CUDA cores, as shown in Fig. \ref{fig1}. As an example, the NVIDIA P100 GPU-accelerators \cite{nvidiap100} have 56 SMs and each SM has 64 CUDA cores with a total of 3584 FP32 cores/GPU or 1792 FP64 cores/GPU, while the NVIDIA V100 \cite{nvidiav100} has 80 SMs and each SM has 64 CUDA cores with a total of 5120 FP32/GPU or 2560 FP64/GPU, where FP32 and FP64 correspond to the single-precision Floating Point (FP) (i.e. 32 bit) and to the double precision (64 bit), respectively. 

\begin{figure}[h!]
\centering
\includegraphics[width=16.cm,height=12cm]{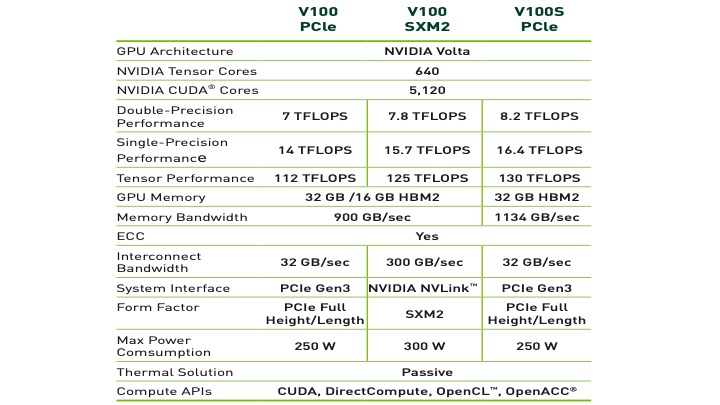}
\caption{\label{fig2} Specification of the architecture of the NVIDIA Volta GPU taken from the ref. \cite{nvidiav100}.}
\end{figure}

Various NVIDIA GPU-architectures \cite{nvidiaarch} exist. As an example, we present in Fig. \ref{fig2} the characteristic of the NVIDIA V100 Volta architecture. As shown in the figure, the peak performance of the NVIDIA Volta depends on the specified architecture: V100 PCle, V100 SXMe and V100S PCle, which in turn depends, in particular, on the  Memory Bandwidth. For instance, the double precision performance associated with each architecture is respectively 7, 7.8 and 8.2 TFLOPS (or TeraFLOPS). Here 1 TFLOPS= $10^{12}$ calculations per second, where FLOPS (Floiting-Point of Opertaions Per Second), in general, defines a measure of the speed of a computer to perform arithmetic operations. The peak performance can be calculated theoretically based on the following expression for a single processor
\begin{equation}\label{flops}
FLOPS = (Clock speed) \times (cores) \times \frac{FLOP}{cycle},  
\end{equation}
 where FLOP is a way of encoding real numbers (i.e. FP64 or FP32 or ...). One can check the validity of the expression by calculating, for instance, the peak performance for V100 PCle, in which the Clock speed (or GPU Boost Clock) is 1.38 GHz \cite{archv100detail}. The total FLOPS is (1.38 $10^{9}$ cycle/second)x5120xFLOP/cycle, which is 7.065 $10^{12}$ FLOP per second or also 7.065 TFLOPS in accordance with the peak performance indicated in Fig. \ref{fig2}.

\subsection{Experiment on OpenACC offloading}\label{acc}

\begin{figure}[h!]
\centering
\includegraphics[width=16.cm,height=10cm]{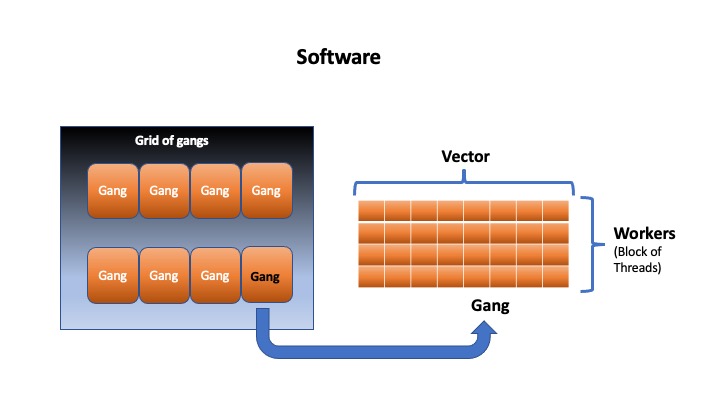}
\caption{\label{fig3} Schematic representation of the concept of parallelism (see text for more details).}
\end{figure}

We begin first by illustrating the functionality of the OpenACC model \cite{accmodel} in terms of parallelism, which is specified by the directives \textbf{kernels} or \textbf{parallel loop}. The concept of parallelism is defined precisely by the generic directives: \textbf{gang}, \textbf{worker} and \textbf{vector} as schematically depicted in Fig. \ref{fig3}. Here, the compiler initiates the parallelism by generating parallel gangs, in which each gang consists of a set of workers represented by a matrix of threads as depicted in the inset of Fig. \ref{fig3}. This group of threads within a gang executes the same instruction (SIMT, Single Instruction Multiple Threads) via a vectorization process. In other words, a block of loops is assigned to each gang, which gets vectorized and executed by a group of threads. Specifically, each thread executes the same kernel program but operates on different parts of the offloaded region. 

By combining the two pictures displayed in Figs. \ref{fig1} and \ref{fig2}, one can say that the execution of the parallelism, which is specified by the \textbf{parallel loop} construct, is mapped on the GPU-device in the following way: each streaming multiprocessor is associated to one gang of threads generated by the directive \textbf{gang}, in which a block of loops is assigned to. In addition, this block of loops is run in parallel on the CUDA cores via the directive \textbf{vector}. The description of these directives and others implemented in our OpenACC mini-application is summarized in the Table \ref{table1}.  

\begin{figure}[h!]
\centering
\includegraphics[width=12.cm,height=8cm]{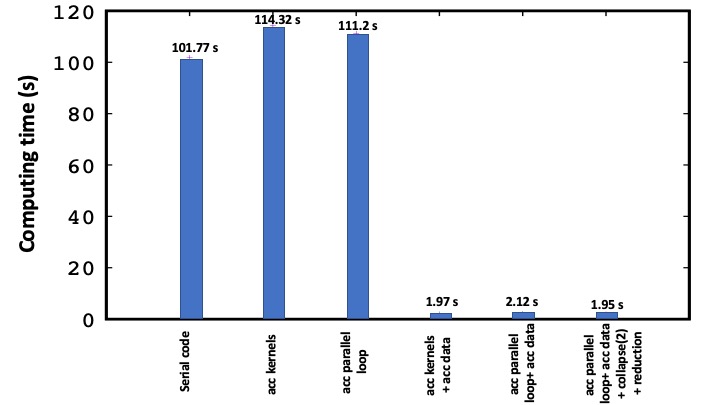}
\caption{\label{fig4} Performance of different OpenACC directives.}
\end{figure}

We move now to discuss our OpenACC experiment, in which we evaluate the performance of different compute constructs and clauses and interpret their role. The OpenACC-based code is shown below. In the first mini-application, only the directive \textbf{parallel loop} is introduced. Here the construct \textbf{parallel} indicates that the compiler will generate a number of parallel gangs to execute the compute region redundantly. When it is combined with the clause \textbf{loop}, the compiler will perform the parallelism over all the generated gangs\footnote{Different gangs operate independently.} for the offloaded region. In this case the compiler copies the data first to a device in the beginning of the loop and then copies it back to the host at the end of the loop. This process repeats itself at each iteration, which makes it time consuming, thus rending the GPU-acceleration inefficient. This inefficiency is shown in Fig. \ref{fig4} and manifests by the increase of the computing time: 111.2 s compared to 101.77 s in the serial case. This low performance is also observed when using the construct \textbf{kernels}.   

To overcome this issue, one needs to copy the data to a device only in the beginning of the iteration and to copy them back to the host at the end of the iteration, once the result converges. This can be done by introducing the data locality concepts via the directives \textbf{data}, \textbf{copyin} and \textbf{copyout}, as shown in the second mini-application, which is referred to as 'OpenACC with data locality'. Here, the clause \textbf{copyin} transfers the data to a GPU-device, while the clause \textbf{copyout} copies the data back to the host. Implementing this approach shows a vast improvement of the performance: the computing time gets reduced by almost a factor of 53: it decreases from 111.2 s to 2.12 s. One can further tune the process by adding additional control, for instance, by introducing the \textbf{collapse} clause. Collapsing two or more loops into a single loop is beneficial for the compiler, as it allows to enhance the parallelism when mapping the compute region into a device. In addition, one can specify the clause \textbf{reduction}, which allows to compute the maximum of two elements in a parallel way. These additional clauses affect slightly the computing time: it goes from 2.12 s to 1.95 s.

For completeness, we compare in Fig. \ref{fig4} the performance of the compute constructs\footnote{When incorporating the constructs \textbf{kernels} or \textbf{parallel loop}, the compiler will generate arrays that will be copied back and forth between the host and the device if they are not already present in the device.} \textbf{kernels} and \textbf{parallel loop}. These directives tell the compiler to transfer the control of a compute region to a GPU-device and execute it in a sequence of operations. Although these two constructs have a similar role, they differ in terms of mapping the parallelism into a device. Here, when specifying the \textbf{kernels} construct, the compiler performs the partition of the parallelism explicitly by choosing the optimal numbers of gangs, workers and the length of the vectors and also some additional clauses. Whereas, the use of the \textbf{parallel loop} construct offers some additional functionality: it allows the programmer to control the execution in a device by specifying additional clauses. At the end, the performance remains roughly the same as shown in Fig. \ref{fig4}: the computing time is 1.97 s for the \textbf{kernels} directive and 1.95 s for the \textbf{parallel loop} directive. 

\begin{lstlisting}[language=Fortran]
!OpenACC without including data locality
do while (max_err.gt.error.and.iter.le.max_iter) 
!$acc parallel loop gang worker vector
   do j=2,ny-1 
      do i=2,nx-1 
         df_x = f(i+1,j) + f(i-1,j) 
         df_y = f(i,j+1) + f(i,j-1) 
         f_k(i,j) = 0.25*(df_x + df_y) 
      enddo 
    enddo  
!$acc end parallel     
    max_err=0. 

!$acc parallel loop
    do j=2,ny-1 
       do i=2,nx-1 
          max_err = max(dabs(f_k(i,j) - f(i,j)),max_err) 
          f(i,j) = f_k(i,j) 
       enddo 
    enddo  
!$acc end parallel
    iter = iter +1 
enddo 
\end{lstlisting}
 
\begin{lstlisting}[language=Fortran]
!OpenACC with data locality
!$acc data copyin(f) copyout(f_k)
do while (max_err.gt.error.and.iter.le.max_iter) 
!$acc parallel loop gang worker vector collapse(2)
   do j=2,ny-1 
      do i=2,nx-1 
         df_x = f(i+1,j) + f(i-1,j) 
         df_y = f(i,j+1) + f(i,j-1) 
         f_k(i,j) = 0.25*(df_x + df_y) 
      enddo 
    enddo  
!$acc end parallel     
    max_err=0. 

!$acc parallel loop collapse(2) reduction(max:max_err)  
    do j=2,ny-1 
       do i=2,nx-1 
          max_err = max(dabs(f_k(i,j) - f(i,j)),max_err) 
          f(i,j) = f_k(i,j) 
       enddo 
    enddo  
!$acc end parallel    
    iter = iter +1 
enddo 
!$acc end data
\end{lstlisting}

\subsubsection{Compiling and running OpenACC-program}\label{compilingacc}

We run our OpenACC-program on the NVIDIA-GPU P100. The syntax of the compilation process is
\begin{lstlisting}
$ nvfortran -fast -acc -Minfo=accel -o laplace_acc.exe laplace_acc.f90
or
$ nvfortran -gpu=tesla:cc60 -Minfo=accel -o laplace_acc.exe laplace_acc.f90
\end{lstlisting}
where the flags \textit{-acc} and \textit{-gpu=[target]} enables OpenACC directives. The option \textit{[target]} reflects the name of the GPU device. The latter is set to be \textit{[tesla:cc60]} for the device name Tesla P100 and \textit{[tesla:cc70]} for the tesla V100 device. This information can be viewed by running the command \textbf{pgaccelinfo}. Last, the flag option \textit{-Minfo} enables the compiler to print out the feedback messages on optimizations and transformations.

The generated binary (i.e. \textbf{laplace\_acc.exe}) can be launched with the use of a Slurm script as follows

\begin{lstlisting}[language=bash]
#!/bin/bash
#SBATCH --account=<project-account> 
#SBATCH --job-name=laplace_acc
#SBATCH --partition=accel --gpus=1
#SBATCH --qos=devel
#SBATCH --time=00:01:00
#SBATCH --mem-per-cpu=2G
#SBATCH -o laplace_acc.out

#loading modules
module purge
module load NVHPC/21.2
 
$ srun ./laplace_acc.exe
\end{lstlisting}

In the script above, the option \textit{--partition=accel} enables the access to a GPU device connected to a node. One can also use the command \textit{sinfo} to get information about which nodes are connected to the GPUs. 

\begin{lstlisting} 
The compilation process requires loading a NVHPC module, e.g. NVHPC/21.2 or another version.
\end{lstlisting}

\subsection{Experiment on OpenMP offloading}\label{omp}

In this section, we carry out an experiment on OpenMP \cite{ompmodel} offloading by adopting the same scenario as in the previous section \ref{acc} but with the use of a different GPU-architecture: AMD Mi100 accelerator. The functionality of OpenMP is similar to the one of OpenACC, although the terminology is different [cf. Fig. \ref{fig1}]. In the OpenMP concept, a block of loops is offloaded to a device via the construct \textbf{target}. A set of threads is then created on each compute unit (CU) (analogous to a streaming multiprocessor in NVIDIA terminology) [cf. Fig. \ref{fig1}] by means of the directive \textbf{teams} to execute the offloaded region. Here, the offloaded region (e.g. a block of loops) gets assigned to teams via the clause \textbf{distribute}, and gets executed on the processing elements (PEs) (analogous to CUDA cores) by means of the directive \textbf{parallel do simd}. These directives define the concept of parallelism in OpenMP offloading. 

\begin{figure}[h!]
\centering
\includegraphics[width=12.cm,height=8cm]{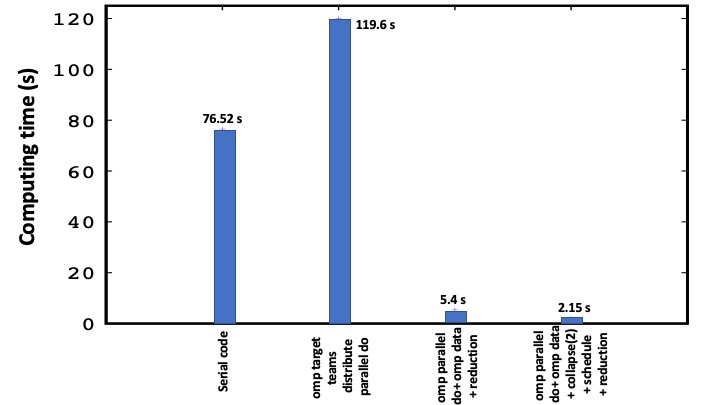}
\caption{\label{fig5} Performance of different OpenMP directives.}
\end{figure}

The concept of parallelism is implemented using the same model described in Section \ref{model}. The implementation is presented below for two cases: (i) OpenMP without introducing the data directive and (ii) OpenMP with the data directive. This Comparison allows us to identify the benefit of data management during the data-transfer between the host and a device. This in turn provides some insights into the performance of the OpenMP offload features. In the the OpenMP mini-application (i), the arrays \textbf{f} and \textbf{f$_k$}, which define the main components of the compute region, are copied from the host to a device and back, respectively via the clause \textbf{map}. Note that specifying the \textbf{map} clause in this case is optional. Once the data are offloaded to a device, the parallelism gets executed according to the scenario described above. This scheme repeats itself at each iteration, which causes a low performance as shown in Fig. \ref{fig5}. Here the computing time is 119.6 s, which is too high compared to 76.52 s in the serial case. A similar behavior is observed in the OpenACC mini-application.

The OpenMP performance, however is found to be improved when introducing the directive \textbf{data} in the beginning of the iteration (see the code 'OpenMP including data directives'). This implementation has the advantage of keeping the data in the device during the iteration process and copying them back to the host only at the end of the iteration. By doing so, the performance is improved by almost a factor of 22, as depicted in Fig. \ref{fig5}: it goes from 119.6 s in the absence of the data directive to 5.4 s when the directive is introduced. As in the OpenACC mini-application, the performance can be further tuned by introducing additional clauses, specifically, the clauses \textbf{collapse} and \textbf{schedule} which are found to reduce the computing time from 5.4 s to 2.15 s. 

The description of the compute constructs and clauses used in our OpenMP mini-application is provided in the Table. \ref{table1} together with those of OpenACC. 

\begin{lstlisting}[language=Fortran]
!OpenMP without including data directives
do while (max_err.gt.error.and.iter.le.max_iter) 
!$omp target teams distribute parallel do simd map(to:f) map(from:f_k) 
   do j=2,ny-1 
      do i=2,nx-1 
         df_x = f(i+1,j) + f(i-1,j) 
         df_y = f(i,j+1) + f(i,j-1) 
         f_k(i,j) = 0.25*(df_x + df_y) 
      enddo 
    enddo  
!$omp end target teams distribute parallel do simd     
    max_err=0. 

!$omp target teams distribute parallel do simd
    do j=2,ny-1 
       do i=2,nx-1 
          max_err = max(dabs(f_k(i,j) - f(i,j)),max_err) 
          f(i,j) = f_k(i,j) 
       enddo 
    enddo  
!$omp end target teams distribute parallel do simd
    iter = iter +1 
enddo 
\end{lstlisting}

\begin{lstlisting}[language=Fortran]
!OpenMP including data directives
!$omp target data map(to:f) map(from:f_k)
do while (max_err.gt.error.and.iter.le.max_iter) 
!$omp target teams distribute parallel do simd collapse(2) schedule(static,1)  
   do j=2,ny-1 
      do i=2,nx-1 
         df_x = f(i+1,j) + f(i-1,j) 
         df_y = f(i,j+1) + f(i,j-1) 
         f_k(i,j) = 0.25*(df_x + df_y) 
      enddo 
    enddo  
!$omp end target teams distribute parallel do simd     
    max_err=0. 

!$omp target teams distribute parallel do simd collapse(2) schedule(static,1) reduction(max:max_err)
    do j=2,ny-1 
       do i=2,nx-1 
          max_err = max(dabs(f_k(i,j) - f(i,j)),max_err) 
          f(i,j) = f_k(i,j) 
       enddo 
    enddo  
!$omp end target teams distribute parallel do simd
    iter = iter +1 
enddo 
!$omp end target data
\end{lstlisting}

\subsubsection{Compiling and running OpenMP-program}\label{compilingomp}

Our OpenMP benchmark test runs on AMD Mi100 accelerator. The syntax of the compilation process can be written in the following form:

\begin{lstlisting}
flang -fopenmp=libomp -fopenmp-targets=<target> -Xopenmp-target=<target> -march=<arch> laplace_omp.f90
\end{lstlisting}

The flag \textit{-fopenmp} activates the OpenMP directives. The option \textit{-fopenmp-targets=$<$target$>$} is used to enable the target offloading to GPU-accelerators and tells the \textit{Flang} compiler to use \textit{$<$target$>$=amdgcn-amd-amdhsa} as the AMD target. The \textit{-Xopenmp-target} flag enables options to be passed to the target offloading toolchain. In addition, we need to specify the architecture of the GPU to be used. This is done via the flag \textit{-march=$<$arch$>$}, where \textit{$<$arch$>$} specifies the name of the GPU-architecture. This characteristic feature can be extracted from the machine via the command \textbf{rocminfo}. For instance, the AMD Mi100 accelerator architecture is specified by the flag \textit{-march=gfx908 amd-arch}. 

\begin{lstlisting}
The compilation process requires loading a AOMP module, e.g. AOMP/13.0-2-GCCcore-10.2.0 or a newer version.
\end{lstlisting}

\begin{table}[h!]
\centering
\begin{tabular}{|p{24em}|p{24em}|}
\hline
OpenACC & OpenMP \\
\hline
\hline
\begin{lstlisting}[language=Fortran]
!$acc data copyin(f) copyout(f_k) 
do while (max_err.gt.error.and.iter.le.max_iter) 
!$acc parallel loop gang worker vector collapse(2) 
   do j=2,ny-1  
      do i=2,nx-1  
         df_x = f(i+1,j) + f(i-1,j) 
         df_y = f(i,j+1) + f(i,j-1) 
         f_k(i,j) = 0.25*(df_x + df_y) 
      enddo  
    enddo   
!$acc end parallel    

    max_err=0.  

!$acc parallel loop collapse(2) reduction(max:max_err) 

    do j=2,ny-1
       do i=2,nx-1  
          max_err = max(dabs(f_k(i,j) - f(i,j)),max_err) 
          f(i,j) = f_k(i,j) 
       enddo 
    enddo  
!$acc end parallel 

    iter = iter +1  
enddo  
!$acc end data 
\end{lstlisting}
& 
\begin{lstlisting}[language=Fortran]
!$omp target data map(to:f) map(from:f_k)
do while (max_err.gt.error.and.iter.le.max_iter) 
!$omp target teams distribute parallel do simd collapse(2) schedule(static,1)  
   do j=2,ny-1 
      do i=2,nx-1 
         df_x = f(i+1,j) + f(i-1,j) 
         df_y = f(i,j+1) + f(i,j-1) 
         f_k(i,j) = 0.25*(df_x + df_y) 
      enddo 
    enddo  
!$omp end target teams distribute parallel do simd     
    max_err=0. 

!$omp target teams distribute parallel do simd collapse(2) schedule(static,1) reduction(max:max_err)
    do j=2,ny-1 
       do i=2,nx-1 
          max_err = max(dabs(f_k(i,j) - f(i,j)),max_err) 
          f(i,j) = f_k(i,j) 
       enddo 
    enddo  
!$omp end target teams distribute parallel do simd
    iter = iter +1 
enddo 
!$omp end target data
\end{lstlisting}
\\
\hline
\end{tabular}
\caption{Fortran Mini-application: OpenACC (left-hand-side) versus OpenMP (right-hand-side).}
\label{table0}
\end{table}

\subsection{Mapping OpenACC to OpenMP}\label{mapping}

In this section, we present a direct comparison between the OpenACC and OpenMP offload features. This comparison is illustrated in the compute code in Table \ref{table0}. A closer look at OpenACC and OpenMP codes reveals some similarities and differences in terms of constructs and clauses. The meaning of these directives are summarized in Table \ref{table1}. Here, evaluating the behavior of OpenACC and OpenMP by one-to-one mapping is a key feature for an effort of porting OpenACC to OpenMP offloading. Based on this comparison, it is seen that the syntax of both programming models is so similar, thus making the implementation of the translation procedure at the syntactic level straightforward. Therefore, carrying out such a comparison is critical for determining the correct mappings to OpenMP.

\begin{table}[h!]
\centering
\begin{tabular}{|p{10em}|p{12em}|p{26em}|}
\hline
OpenACC & OpenMP & interpretation \\
\hline
\hline
acc parallel & omp target teams & to execute a compute region on a device \\
acc kernels  & No explicit counterpart   & - -  \\
acc parallel loop gang worker vector & omp target teams distribute parallel do simd & to parallelize a block of loops on a device \\
acc data     & omp target data & to share data between multiple parallel regions in a device \\
\hline
acc loop & omp teams distribute & to workshare for parallelism on a device \\
acc loop gang & omp teams(num\_teams) & to partition a loop across gangs/teams \\
acc loop worker & omp parallel simd & to partition a loop across threads \\
acc loop vector & omp parallel simd & - -  \\
num\_gangs       & num\_teams         & to control how many gangs/teams are created \\
num\_workers     & num\_threads       & to control how many worker/threads are created in each gang/teams \\
vector\_length   & No counterpart    & to control how many data elements can be operated on \\
\hline
acc create() & omp map(alloc:) & to allocate a memory for an array in a device \\
acc copy()   & omp map(tofrom:) & to copy arrays from the host to a device and back to the host \\
acc copyin() & omp map(to:) & to copy arrays to a device \\
acc copyout() & omp map(from:) & to copy arrays from a device to the host \\
\hline
acc reduction(operator:var) & omp reduction(operator:var) & to reduce the number of elements in an array to one value \\
acc collapse(N)  & omp collapse(N)   & to collapse N nested loops into one loop \\
No counterpart  & omp schedule(,)  & to schedule the work for each thread according to the collapsed loops \\
private(var)         & private(var)          & to allocate a copy of the variable $var$ on each gang/teams \\
firstprivate    & firstprivate     & to allocate a copy of the variable `var` on each gang/teams and to initialise it with the value of the local thread \\

\hline
\end{tabular}
\caption{Description of various directives and clauses: OpenACC vs OpenMP.}
\label{table1}
\end{table}
 
Note that further description about directives is provided in Refs. \cite{accguide,ompguide} and details about library routines can be found in Ref. \cite{libacc} for OpenACC and in Ref. \cite{libomp} for OpenMP.

\section{Open-source OpenACC compilers}\label{clacccompiler}

 For completeness, we provide in this section some highlights of the available open-source OpenACC compilers. According to the work of J. Vetter et \textit{al.} \cite{ieee} and the OpenACC website \cite{accweb}, the only open-source compiler that supports OpenACC offloading to NVIDIA and AMD accelerators is GCC 10. Recently, there has been an effort in developing an open-source compiler to complement the existing one, thus allowing to perform experiments on a broad range of architectures. The compiler is called Clacc \cite{ieee} and its development is funded by the Exascale Computing Project \cite{project} and is further described by the work of J. Vetter et \textit{al.} \cite{ieee}. We thus focus here on providing some basic features of the Clacc compiler platform, without addressing deeply the fundamental aspect of the compiler, which is beyond the scope of this documentation..

Clacc is an open-source OpenACC compiler platform that has support for Clang \cite{clang} and LLVM \cite{llvm}, and aims at facilitating GPU-programming in its broad use. The key behind the design of Clacc is based on translating OpenACC to OpenMP, taking advantage of the existing OpenMP debugging tools to be re-used for OpenACC. Clacc was designed to mimic the exact behavior of OpenMP as explicit as possible. The Clacc strategy for interpreting OpenACC is based on one-to-one mapping of OpenACC directives to OpenMP directives \cite{ieee} as we have already shown in the Table \ref{table1} above. 

Despite the new development of Clacc compiler platform, there is still major need to further extend the compiler as it suffers from some limitations, mainly \cite{ieee}: (i) in the Clacc's design, translating OpenACC to OpenMP in Clang is currently supported only in C but not yet in C++. (ii) Clacc has so far focused primarily on compute constructs, and thus lacks support of data-sharing between the CPU-host and a GPU-device. These limitations however are expected to be overcome in the near future. So far, Clacc has been tested and benchmarked against a series of different configurations, and it is found to provide an acceptable GPU-performance, as stated in Ref. \cite{project}. Note that Clacc is publicly available in Ref. \cite{github}.

\section{Conclusion}\label{conclusion}

In conclusion, we have presented an overview of the GPU-architecture as well as the OpenACC and OpenMP offload features via an application based on solving the Laplace equation in a 2D uniform grid. This benchmark application was used to experiment the performance of some of the basic directives and clauses in order to highlight the gain behind the use of GPU-accelerators. The performance here was found to be improved by almost a factor of 52. We have also presented an evaluation of differences and similarities between OpenACC and OpenMP programming models. Furthermore, we have illustrated a one-to-one mapping of OpenACC directives to OpenMP directives in the aim of porting OpenACC to OpenMP. In this context, we have emphasized the recent development of the Clacc compiler platform, which is an open-source OpenACC compiler, although the platform support is so far limited to C and lacks data-transfer in host-device. 

Last but not least, writing an efficient GPU-based program requires some basic knowledge of the GPU architecture and how regions of a program is mapped into a target device. This documentation thus was designed to provide such basic knowledge in the aim of triggering the interest of developers/users to GPU-programming. 

\section*{Acknowledgments}
The author would like to thank dr. J{\o}rgen Nordmoen for reviewing the manuscript and the GPU-team within the NRIS (Norwegian Research Infrastructure Services) for fruitful discussion. This work used resources provided by UNINETT Sigma2 the National Infrastructure for High Performance Computing and Data Storage in Norway.

\end{document}